# MECHANICAL PROPERTIES OF $CaFe_2As_2$ *AB-INITIO* STUDY

M.V. Samuel[1] C.O. Otieno[1]

1. Department of Physics, Kisii University, P.O BOX 408, Kisii, Kenya.

E-mail:mochamavic@gmail.com

## ABSTRACT

We report results on the *ab-initio* study of the mechanical, electronic and structural properties of the iron Pnictide compound $CaFe_2As_2$ at zero pressure. Ground State energy computation was done within the Density Functional Theory (DFT) using the Projector Augmented Wave (PAW) Pseudo Potentials and the Plane Wave (PW) basis set. The Generalized Gradient Approximation (GGA) facilitated the exchange-correlation. QUANTUM ESPRESSO (QE) code played a crucial role in this study together with THERMO_PW in the evaluation of these mechanical properties. From the non-zero positive elastic constants we concluded that the Iron Pnictide compound is mechanically stable. Poisson's ratio values confirms that this compound is brittle and anisotropic. Electronic structure calculations reveals that this compound is a metal.

# Introduction

The revelation of iron Pnictide materials superconducting at high temperatures has renewed research on superconductivity mechanisms[1]. The superconducting iron Pnictide class of compounds(122) have simple structures unlike the 111 family of Pnictides[2].

$AFe_2As_2$ is stable with divalent (A=Ba, Ca, Sr and Eu) atoms. They crystallize at room temperature in $ThCr_2Si_2$ type tetragonal structure[3] with a space group of 14/mmm[4, 5]. $CaFe_2As_2$ transits from tetragonal to orthorhombic and orders antiferromagnetically in the orthorhombic structure[6], indicating stripe like magnetic order[5].

Superconductivity can be achieved by doping and by application of pressure on the compound[7]. Magnetic and structural instabilities are known to be suppressed by electron or hole doping inducing superconductivity. Application of pressure in the parent compounds suppresses magnetic order[8] inducing superconductivity also. $CaFe_2As_2$ contain fragile magnetism active in response to external strain facilitating superconductivity greatly[8, 9]. The material has a structural phase transition to a collapsed tetragonal phase whereby its $c$ parameter is decreased by 10%.[10]

The 3d orbitals in this compound facilitates the low-lying electronic state comprised of many holes and electron bands giving rise to superconductivity and electron pairing[11]. The onset of superconductivity and the appearance of several phases at low temperatures in $CaFe_2As_2$ are extremely sensitive to pressure conditions. In collapsed tetragonal shape, the electron correlation and superconductivity disappear[12]. $CaFe_2As_2$ is transformed to orthorhombic at low temperature[13]. Pressur application by a non-hydrostatic medium makes $CaFe_2As_2$ a superconductor an indication that small uniaxial pressure parts overrun its superconductivity[14].

Cooling $CaFe_2As_2$ through 170K undergoes first order magnetic phase transition[15]. Superconductivity appears above 40K when the single crystal of the compound has been electron doped[16]. This compound contains oxide blocks on the $c$-axis and two dimensional iron (Fe) and arsenic (As) tetrahedron layers which act as charge reservoirs.

The magnetic state of iron in this compound dictates its superconductivity upon which adjustments can be done in turning it on and off through change of order and the fluctuations or existence of iron magnetism[17].

Na-substitution in between $Ca$ and $Fe$ with composition $Ca_5Na_5Fe_2As_2$ induce superconductivity at 20K. Upon application of pressure its restivity lacks any sign of superconductivity but has a reduction of its unit cell volume[18]. Reduction of its unit cell volume and hole doping are very essential in inducing superconductivity. In this paper we present well elaborated First Principle investigations on electronic, structural and mechanical properties of $CaFe_2As_2$ at ground state temperature.

**Computational Method**

We employed the use of Density Functional Theory and QUANTUM ESPRESSO package for the determination of band structure and Density of States (DOS). Also, THERMO_PW incorporated in QE was used in computing the mechanical properties. We also used Projector Augmented Wave (PAW) method which is a technique used in *ab-initio* electronic structure calculation[20]. Pseudo Potential and linear Augmented-Plane-Wave methods perform greater computational calculation with efficiency and accuracy within the scope [21].

PAW was effective because it offered good computational scalability and systematic convergence properties.

We are employing First Principles study involving computation of electronic orbitals. The electrons obey Fermi statistics with the Pauli Exclusion Principle. Born-Oppenheimer approximation which, treats nuclear and electronic structure calculation for a fixed configuration is facilitated. Electronic Schrödinger equation is not solved fully because of its complexity when involving many electrons, it is transformed into an algebraic equation and solved by numerical methods[22].

Hartree-Fock approximation assumes each electron moves in an average potential without consideration of their positions and applies it in the solution of the Schrödinger equation for atoms and electrons. Post-Hartree-Fork electron correlation takes the instantaneous correlation of the electrons into account[23].

**Results and Discussion**

In this section, we report the results obtained from the computational study.

We optimized the cut off energy and k-points and convergence was achieved at 45Ry and 5Bohr respectively as shown in the figures below;

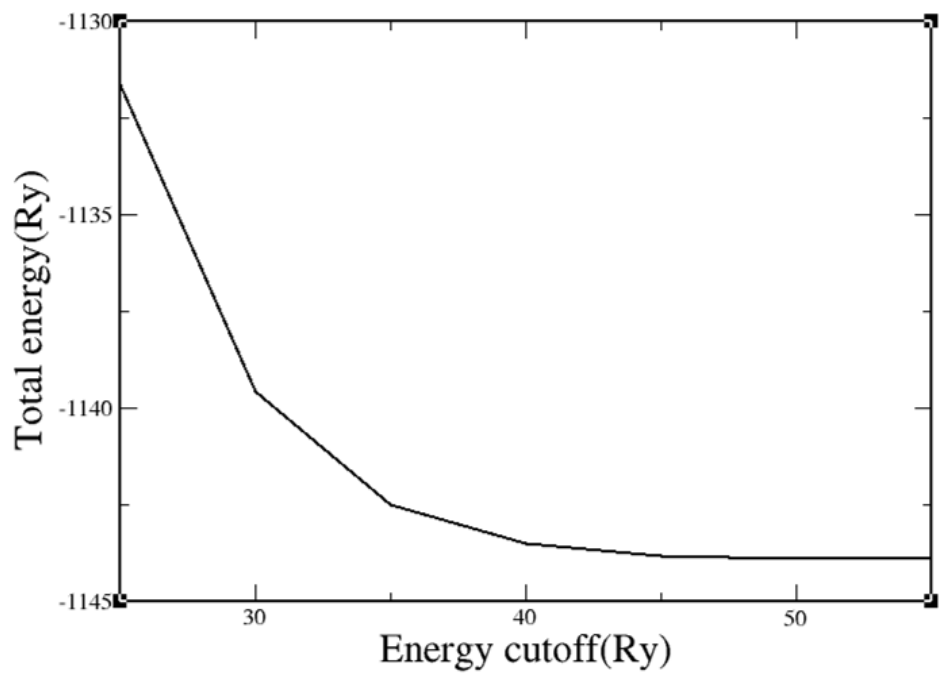

Figure 1: Optimization curve for the energy cut off. Convergence was achieved at 45Ry

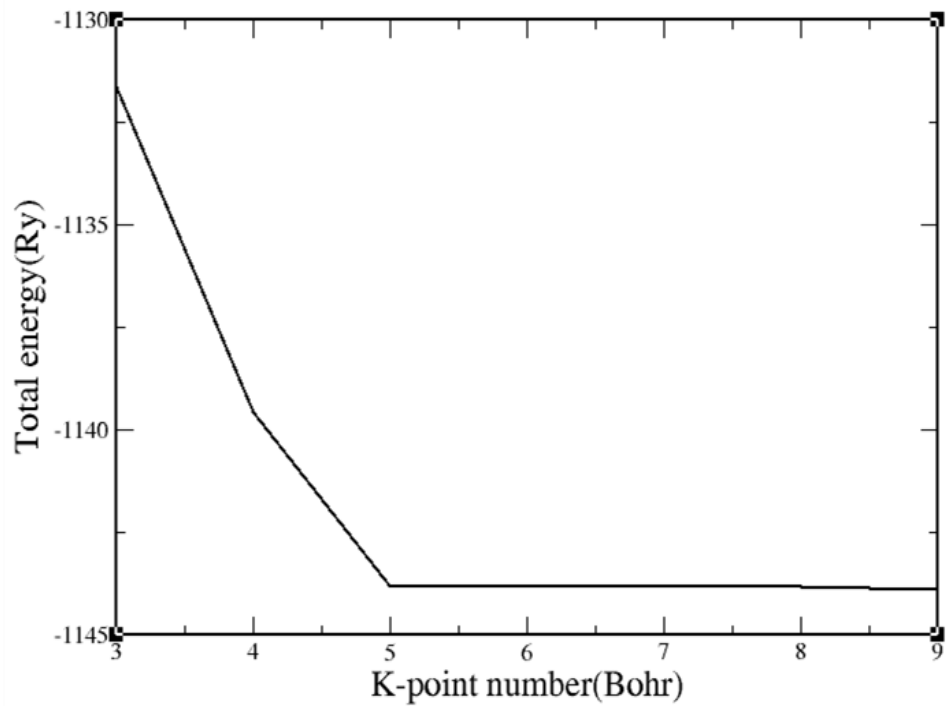

Figure 2: Optimization curve for the k-point. Convergence was achieved at k-point 5.

## Structural properties

$CaFe_2As_2$ at stability has a tetragonal phase with its unit cell consisting of two $Ca$ atoms, two $Fe$ atoms and two $As$ atoms. The stable crystal structure is as shown below;

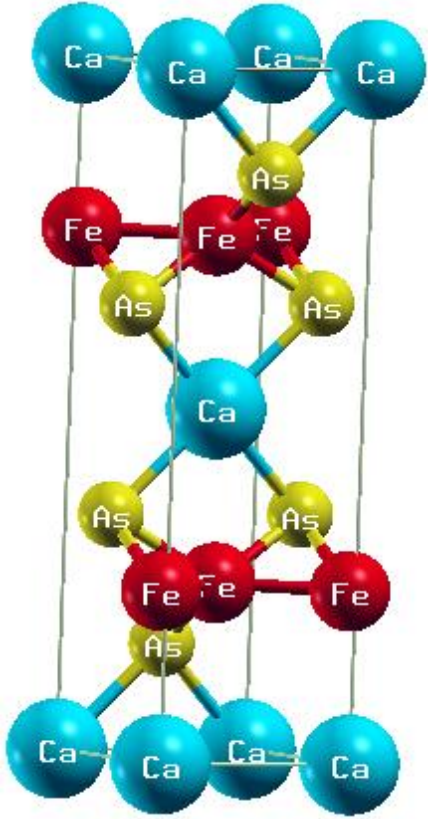

**Fig 3: Crystal structure of $CaFe_2As_2$ drawn by quantum espresso Xcrysden package.**

We did optimization of lattice constants and cut off energy that gave us a relaxed structure free from stress as shown in the table below[3, 7]

*Table 1: Comparison of experimental and theoretical cell dimensions.*

| Parameter | This work | Experimental | Reference |
|---|---|---|---|
| $a_0=b_0$ (ang) | 3.249 | 3.887 | [3] |
| $c_0$ (ang) | 7.493 | 7.5898 | [3] |

The optimized cell parameters have a very small margin of error hence in good agreement with both experimental and theoretical work.

The tetragonal crystal structure has six elastic constants as recorded in the table below[7].

Table 2: Elastic constants of $CaFe_2As_2$

| $C_{ij}$ | Value(GPa) |
|---|---|
| $C_{11}$ | 88.86 |
| $C_{12}$ | 22.58 |
| $C_{13}$ | 28.63 |
| $C_{33}$ | 63.51 |
| $C_{44}$ | 25.95 |
| $C_{66}$ | 31.73 |

$C_{11}$ and $C_{33}$ elastic constants indicate the response of the uniaxial strain. $C_{12}$, $C_{13}$ and $C_{44}$ are related to responses that are shear dominated. $C_{11}$>$C_{33}$ a clear indication that the crystal is more rigid on [1 0 0] and [0 1 0] as compared to that along [0 0 1] plane under uniaxial stress[17]. This reveal about the strength of bonds in the individual planes of the material. It is clear from the Table above that $C_{66}$>$C_{44.}$ This implies that [1 0 0] [0 0 1] shear moduli are easier in comparison with [1 0 0] [0 1 0]. This evaluated nature shows clearly the layer features of $CaFe_2As_2$ crystal structure.[17]

State of mechanically stability for tetragonal structure can be well evaluated by Borh-Huang criteria as shown below[24].

**$C_{ii}$>0 (i=1,3, 4, 6)**

**$C_{11}$+$C_{33}$-2$C_{13}$>0**

**2($C_{11}$+$C_{12}$) +$C_{33}$+4$C_{13}$>0**

**$C_{11}$-$C_{12}$>0**

This compound's elastic constants satisfy all of the above mechanical stability conditions. Values of $C_{ij}$ hence can be used in the evaluation of Poisson's and elastic moduli. According to the Voigt approximation criteria, the bulk and shear moduli isotropy can be acquired by linear combination of elastic constants[25]. With a different format, Reuss obtains estimates for bulky and shear moduli isotropy by the use of single crystal elastic constants[26]. Hill confirms that Voigt and Reuss estimates are lower and upper polycrystalline elastic moduli limits, hence the averages became realistic.

$B = \frac{B_V+B_R}{2}$ and G= $\frac{G_V+G_R}{2}$ [17]

Young modulus $E$ and Poisson's ratio *n* in relation to bulky modulus and shear modulus values are tabulated in the Table 3 below[26].

*Table 3: Mechanical properties of Bulk, Shear and Young's modulus and Poisson*

| **Property** | Voigt Approximation | Reuss Approximation | Voigt-Reuss-Hill Average |
|---|---|---|---|
| Bulk modulus(B) | 44.55 | 43.88 | 44.21 |
| Young modulus(E) | 68.39 | 66.40 | 67.40 |
| Shear modulus (G) | 27.48 | 26.60 | 27.04 |
| Poisson ratio(n) | 0.24 | 0.25 | 0.25 |

Shear modulus indicates the strength of the material unlike bulk modulus. $G>B$ hence $CaFe_2As_2$ mechanical failure should be corrected by application of the shear component as B represents resistance to fracture[27]. Plugh's ratio[17] determines how ductile or brittle material is. The high value indicates ductility whereas low value indicates brittle nature of the material. For $\frac{B}{G}$>1.75 indicates ductility otherwise brittle nature. $\frac{B}{G}$ indicate hardness related inversely whereby the smaller the ratio the harder the material is. For our material $\frac{B}{G} < 1.75$ confirms brittleness nature. Poisson's ratio (*n*) assists in the assessment of the mechanical properties of crystalline solids. Its low value indicates stability against shear[28]. Additionally Poisson's ratio reveals nature of interatomic forces whereby it ranges between 0.25 to 0.50 for central force interaction and outside this range for non-central force interaction. According to Poisson ratio materials whose ratio is less than 0.26, the material undergoes brittle failure and above this ratio it undergoes ductile failure. Poisson's ratio also reveals brittle nature of $CaFe_2As_2$. Young modulus evaluates resistance against the compressive or expansive forces. From our Table 3 the shear modulus $E$ is small even smaller than that of $BaPb_2As_2$ indicating that $CaFe_2As_2$ obviously cannot withstand large tensile stress[17].

Cauchy pressure[29] is the difference between $C_{12}$ and $C_{44}$ elastic constants. This parameter reveals more about the elastic response and carge density of solids. Cauchy pressure will indicate ductility or brittleness failure of crystalline solids. A positive or negative Cauchy pressure indicates ductility or brittleness and reveals chemical bonds. Positive value indicates metallic bonds while the negative one indicates covalent bonds. To our study the Cauchy pressure of $CaFe_2As_2$ is negative hence our material is brittle with covalent bonding characteristics.

On determining whether the material is anisotropic, we made use of the following calculation;

$$A^U = \frac{5G_V}{G_R} + \frac{B_V}{B_R} - 6$$

Whereby, if $A^U$=0 the material would be regarded isotropic. With values drawn from the table above, our calculations indicates that the material is anisotropic with a value of 0.1807 which in agreement with the studies of the parent compound $ThCr_2Si_2$ [27]

The electronic structure properties of band structure and Density of States computed at high symmetry points are as shown below.

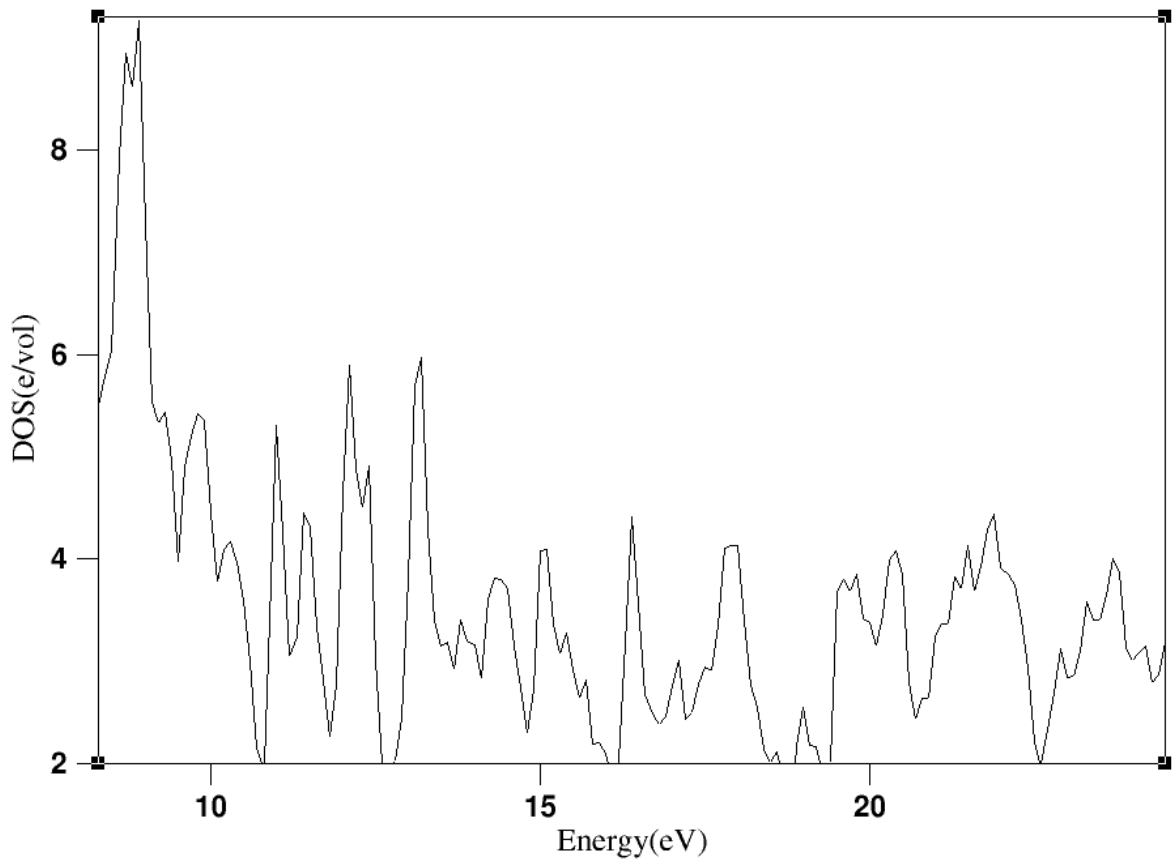

*Fig 4: The Density of states of $CaFe_2As_2$.*

Fe and Ca contain the main weight hence occupies the upper band. Fe dominates the Fermi level at the top of the valence band and contributes greatly in density of state.[30]

$CaFe_2As_2$ Contain energy collective excitations that are evident in the graph, an increase in energy leads the splitting of the peaks. As shown from the graph, vibrations in the x and y planes are rapid indicating the hardening of the phonon[31]. The slope of the graph keeps on reducing an indication of the Phonons approaching extinction [32].

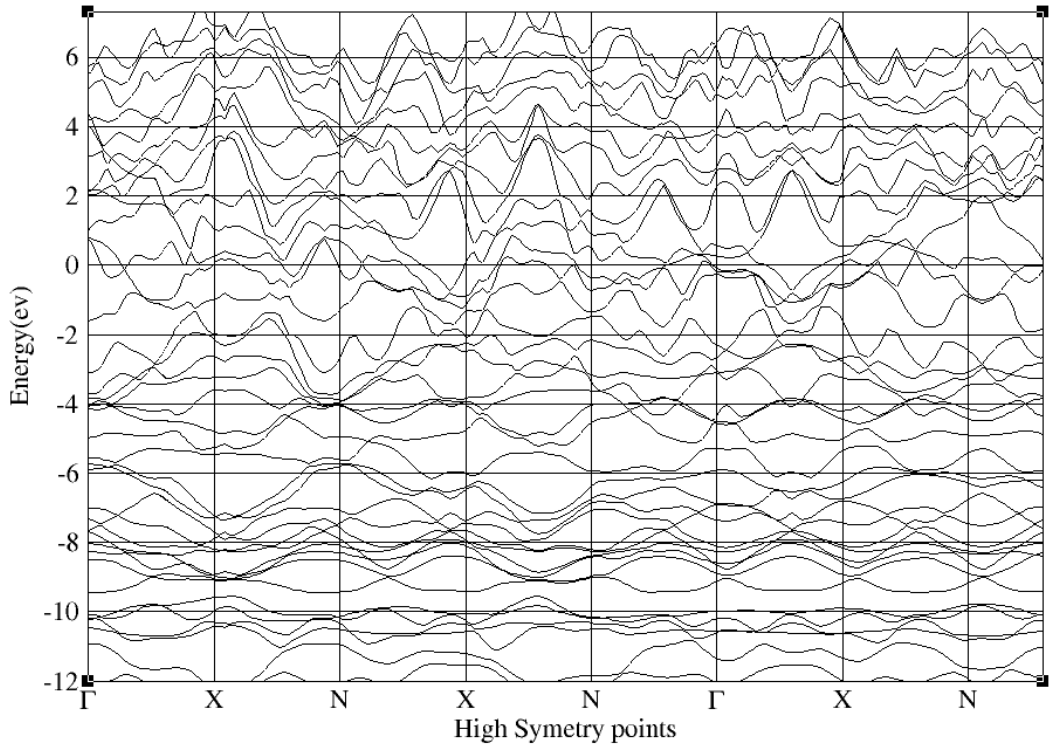

*Fig 5: The band structure*.

The overlapping of the conduction and valence bands is an indication of metallic properties with superconducting gaps in between[33]. The superconducting gaps are moderately anisotropic which puts a powerful restriction on determination of the pairing symmetry. The electrons gain energy through conditions of temperature and pressure crossing over the gaps enabling the material to superconduct. This compound is metallic and comperes well with $ThCr_2As_2$ [34]

## Conclusion

Crystal is rigid on [1 0 0] and [0 1 0] for consideration of uniaxial stresses which indicates the strength of bonds in the individual points of the material. From Poisson ratio in comparison with bulk, shear and young modulus in consideration of interatomic forces confirms that our material is brittle. Convergence was achieved with a minimum energy value of 45eV.

Cauchy pressure is the difference of elastic constants which reveals elastic responses, $CaFe_2As_2$ gives a negative value hence revealing that it is brittle with covalent bonding characteristics. On determining whether our material is anisotropic we employed calculation of $A^U$ using our material's values and indicated that is isotropic which is in good agreement with the parent compound $ThCr_2Si_2$.

From the band structure shown conduction band overlaps with valence band which is a clear indication of the metallic nature of our material.

## **Acknowledgement.**

We acknowledge the resources provided by CHPC for computational work necessary for this work that would not have worked effectively otherwise.

# REFERENCES

1. Dewhurst, J., et al., *First-principles calculation of superconductivity in hole-doped LiBC: $T_c = 65$ K.* Physical Review B, 2003. **68**(2): p. 020504.
2. Samuely, P., et al., *Point contact Andreev reflection spectroscopy of superconducting energy gaps in 122-type family of iron pnictides.* Physica C: Superconductivity, 2009. **469**(9-12): p. 507-511.
3. Ronning, F., et al., *Synthesis and properties of CaFe2As2 single crystals.* Journal of Physics: Condensed Matter, 2008. **20**(32): p. 322201.
4. Soliman, S., *The stacking of antifluorite Fe2P2 layer on the electronic structure of the ternary compounds CaFe2P2, BaFe2P2 and EuFe2P2.* Computational Materials Science, 2016. **122**: p. 177-182.
5. Gonnelli, R., et al., *Point-contact spectroscopy in Co-doped CaFe2As2: nodal superconductivity and topological Fermi surface transition.* Superconductor Science and Technology, 2012. **25**(6): p. 065007.
6. Mittal, R., et al., *Phonon spectra inCaFe2As2andCa0.6Na0.4Fe2As2: Measurement of the pressure and temperature dependence and comparison withab initioand shell model calculations.* Physical Review B, 2009. **79**(14).
7. Omboga, N. and C. Otieno, *Structural and electronic properties of the iron pnictide compound EuFe2As2 from first principles.* Journal of Physics Communications, 2020. **4**(2): p. 025007.
8. Bud'ko, S.L., et al., *Transition to collapsed tetragonal phase in CaFe 2 As 2 single crystals as seen by Fe 57 Mössbauer spectroscopy.* Physical Review B, 2016. **93**(2): p. 024516.
9. Sanna, A., et al., *First-principles study of superconducting Rare-earth doped CaFe2As2.* arXiv preprint arXiv:1406.6513, 2014.
10. Walsh, A., et al., *Structural, magnetic, and electronic properties of the Co-Fe-Al oxide spinel system: Density-functional theory calculations.* Physical review B, 2007. **76**(16): p. 165119.
11. Kasahara, S., et al., *Abrupt recovery of Fermi-liquid transport following the collapse of the c axis in CaFe 2 (As 1− x P x) 2 single crystals.* Physical Review B, 2011. **83**(6): p. 060505.
12. Choi, K.-Y., et al., *Lattice and electronic anomalies of CaFe 2 As 2 studied by Raman spectroscopy.* Physical Review B, 2008. **78**(21): p. 212503.
13. Mishra, S., et al., *Evidence for anomalous structural behavior in CaFe2As2.* arXiv preprint arXiv:1304.0595, 2013.
14. Qi, Y., et al., *Transport properties and anisotropy in rare-earth doped CaFe2As2 single crystals with Tc above 40 K.* Superconductor Science and Technology, 2012. **25**(4): p. 045007.
15. Li, Z., et al., *Magneto-structural coupling and harmonic lattice dynamics in CaFe2As2 probed by Mössbauer spectroscopy.* Journal of Physics: Condensed Matter, 2011. **23**(25): p. 255701.
16. Mihály, G. and P. Beauchene, *Sliding charge density waves without damping: possible Fröhlich superconductivity in blue bronze.* Solid state communications, 1987. **63**(10): p. 911-914.
17. Parvin, F. and S. Naqib, *Structural, elastic, electronic, thermodynamic, and optical properties of layered BaPd2As2 pnictide superconductor: A first principles investigation.* Journal of Alloys and Compounds, 2019. **780**: p. 452-460.

18. Materne, P., et al., *Coexistence of superconductivity and magnetism in Ca 1− x Na x Fe 2 As 2: Universal suppression of the magnetic order parameter in 122 iron pnictides.* Physical Review B, 2015. **92**(13): p. 134511.
19. Kreyssig, A., et al., *Pressure-induced volume-collapsed tetragonal phase of CaFe 2 As 2 as seen via neutron scattering.* Physical Review B, 2008. **78**(18): p. 184517.
20. Blöchl, P.E., C.J. Först, and J. Schimpl, *Projector augmented wave method: ab initio molecular dynamics with full wave functions.* Bulletin of Materials Science, 2003. **26**(1): p. 33-41.
21. Holzwarth, N., et al., *Comparison of the projector augmented-wave, pseudopotential, and linearized augmented-plane-wave formalisms for density-functional calculations of solids.* Physical Review B, 1997. **55**(4): p. 2005.
22. Corey, G.C. and D. Lemoine, *Pseudospectral method for solving the time-dependent Schrödinger equation in spherical coordinates.* The Journal of chemical physics, 1992. **97**(6): p. 4115-4126.
23. Piskunov, S., et al., *Bulk properties and electronic structure of SrTiO3, BaTiO3, PbTiO3 perovskites: an ab initio HF/DFT study.* Computational Materials Science, 2004. **29**(2): p. 165-178.
24. Voigt, W., *Lehrbuch der kristallphysik.* Vol. 962. 1928: Teubner Leipzig.
25. Hadi, M., et al., *Elastic and thermodynamic properties of new (Zr3− xTix) AlC2 MAX-phase solid solutions.* Computational Materials Science, 2017. **137**: p. 318-326.
26. Hadi, M., et al., *Mechanical behavior, bonding nature and defect processes of Mo2ScAlC2: A new ordered MAX phase.* Journal of Alloys and Compounds, 2017. **724**: p. 1167-1175.
27. Shein, I. and A. Ivanovskii, *Structural, elastic, electronic and magnetic properties of ThCr2Si2 from first-principles calculations.* Solid state communications, 2011. **151**(17): p. 1165-1168.
28. Ravindran, P., et al., *Density functional theory for calculation of elastic properties of orthorhombic crystals: application to TiSi 2.* Journal of Applied Physics, 1998. **84**(9): p. 4891-4904.
29. Eberhart, M.E. and T.E. Jones, *Cauchy pressure and the generalized bonding model for nonmagnetic bcc transition metals.* Physical Review B, 2012. **86**(13): p. 134106.
30. Kurmaev, E., et al., *Electronic structure of CaFe2As2: Contribution of itinerant Fe 3d-states to the Fermi Level.* arXiv preprint arXiv:0902.1141, 2009.
31. McQueeney, R., et al., *Anisotropic three-dimensional magnetism in CaFe 2 As 2.* Physical review letters, 2008. **101**(22): p. 227205.
32. Rende, M., et al., *First-principles study on the lattice dynamics and thermodynamics properties of CaFe2As2.* Physica B: Condensed Matter, 2010. **405**(19): p. 4226-4230.
33. Diehl, J., et al., *Correlation effects in the tetragonal and collapsed-tetragonal phase of CaFe 2 As 2.* Physical Review B, 2014. **90**(8): p. 085110.
34. Singh, D.J., *Electronic structure of BaCu 2 As 2 and SrCu 2 As 2: s p-band metals.* Physical Review B, 2009. **79**(15): p. 153102.